\global\def\draftcontrol{0}

   \def\versionno{$U(2)$ on space-time orbifolds}
\catcode`\@=11

\expandafter\ifx\csname draftcontrol\endcsname\relax\global\def\draftcontrol{0} 
\fi 

{\count255=\time\divide\count255 by 60 
\xdef\hourmin{\number\count255} 
\multiply\count255 by-60\advance\count255 by\time 
\xdef\hourmin{\hourmin:\ifnum\count255<10 0\fi\the\count255}} 
\def\draftdate{\number\month/\number\day/\number\year\ \ \ \hourmin } 

\newcommand\makepapertitle{\par

  \begingroup 
    \renewcommand\thefootnote{\@fnsymbol\c@footnote}% 
    \def\@makefnmark{\rlap{\@textsuperscript{\normalfont\@thefnmark}}}% 
    \long\def\@makefntext##1{\parindent 1em\noindent 
            \hb@xt@1.8em{% 
                \hss\@textsuperscript{\normalfont\@thefnmark}}##1}% 
     \newpage 
     \global\@topnum\z@   % Prevents figures from going at top of page. 
     \@makepapertitle 
     \thispagestyle{empty}\@thanks 
  \endgroup 
  \setcounter{footnote}{0}% 
  \global\let\thanks\relax 
  \global\let\makepapertitle\relax 
  \global\let\@makepapertitle\relax 
  \global\let\@thanks\@empty 
  \global\let\@author\@empty 
  \global\let\@date\@empty 
  \global\let\@title\@empty 
  \global\let\title\relax 
  \global\let\author\relax 
  \global\let\date\relax 
  \global\let\and\relax 
  \def\version{\let\version\@version\@gobble} 
} 
\def\@makepapertitle{% 
  \newpage 
   \ifnum\draftcontrol=1 {} 
   \version\versionno 
   \vskip 5.5em% 
   \else 
   \hfill\hbox to 3cm {\parbox{4.5cm}{\@pubnum}\hss}% 
   \vskip 6.5em% 
   \fi 
   \begin{center}% 
   \let \footnote \thanks 
      {\hskip -0\textwidth \hbox to 1\textwidth% 
        {\centerline{\Large\bf{\noindent\@title}}}}% 
     \vskip 2em% 
     {\normalsize%\large 
       \lineskip .5em% 
       \begin{tabular}[t]{c}% 
         \@author 
       \end{tabular}\par}% 
     \vskip 1.5em% 
     {\@bstract}% 
     \end{center}% 
     \vfill
     \@date%
     \vskip 1.5em%
   \par 
} 

\gdef\@pubnum{} 
\def\pubnum#1{% 
  \gdef\@pubnum{#1}} 

\gdef\@bstract{} 
\def\Abstract#1{% 
  \gdef\@bstract{% 
   \parbox{\textwidth-0pc}{% 
   \centerline{\bf Abstract}\penalty1000 
   \noindent%\abstractfont \baselineskip=12pt 
   \renewcommand\baselinestretch{1.0} 
   {#1}}} 
} 

\gdef\@email{}
\def\email#1{%
   \gdef\@email{%
   Email: {\tt #1}}
}

\def\ps@paper{\let\@mkboth\@gobbletwo% 
     \ifnum\draftcontrol=1 
        \def\@oddfoot{\hbox to \textwidth{\tiny \versionno \hfil\tiny\draftdate}% 
        \hskip -\textwidth \hbox to \textwidth{\hfil\rm\thepage\hfil}}% 
     \else\def\@oddfoot{\hbox to \textwidth{\hfil\rm\thepage\hfil}} 
     \fi 
     \let\@evenfoot\@oddfoot 
} 
\def\body{\clearpage 
          \pagestyle{paper} 
        } 
\newenvironment{acknowledgments}{% 
\vskip 3.25ex 
\addcontentsline{toc}{section}{Acknowledgments}
\noindent {\bf Acknowledgments} 
} 

\def\@version#1{\ifnum\draftcontrol=1 
\typeout{}\typeout{#1}\typeout{} 
\vskip3mm\centerline{\hbox{\fbox{\normalsize{\tt DRAFT -- #1 -- } 
                   {\draftdate}}}}\vskip3mm 
\fi} 
\let\version\@version 
\long\def\eqlabel#1{\ifnum\draftcontrol=1 
                    \tag@false  % there are some problems with multline without this 
                    \tag*{(\theequation) \hbox to -0.2cm{\hspace{0cm}\small{#1}\hss}} 
                    \refstepcounter{equation}  
                    \edef\@currentlabel{\theequation} 
                    \ltx@label{#1}          % use old LaTeX \label instead of new definition 
                    \else 
                    \label{#1} 
                    \fi 
                    } 
\let\st@bibitem\@bibitem 
\let\st@lbibitem\@lbibitem 
\ifnum\draftcontrol=1 
  \def\@bibitem#1{% 
    \st@bibitem{#1}\a@@label{#1}\ignorespaces} 
  \def\@lbibitem[#1]#2{% 
    \st@lbibitem[#1]{#2}\a@@label{#2}\ignorespaces} 
  \def\a@@label#1{% 
    \gdef\a@lab{\smash{\normalfont\small#1}} 
    \ifvmode 
      \if@inlabel 
        \global\setbox\@labels\hbox{% 
          \llap{\a@lab\let\a@lab\relax 
                \kern\@totalleftmargin\kern\marginparsep}% 
          \box\@labels}% 
      \fi 
    \fi} 
\fi 
\documentclass[12pt,letterpaper]{article} 
\usepackage{amsmath,bm,amsfonts,amssymb,array,calc,amsthm,rotating}
\usepackage{epsfig,psfrag} 
\usepackage{rotating}
\usepackage{amscd}
\usepackage{graphicx}
\usepackage{color}
\usepackage[colorlinks=false]{hyperref}
\renewcommand\baselinestretch{1.25} 
\renewcommand\section{\@startsection {section}{1}{\z@}% 
                                   {-3.5ex \@plus -1ex \@minus -.2ex}% 
                                   {2.3ex \@plus.2ex}% 
                                   {\normalfont\large\bfseries}} 
\renewcommand\subsection{\@startsection{subsection}{2}{\z@}% 
                                   {-3.25ex\@plus -1ex \@minus -.2ex}% 
                                   {1.5ex \@plus .2ex}% 
                                   {\normalfont\normalsize\bfseries}} 
\renewcommand\subsubsection{\@startsection{subsubsection}{3}{\z@}% 
                                   {-3.25ex\@plus -1ex \@minus -.2ex}% 
                                   {1.5ex \@plus .2ex}% 
                                   {\normalfont\normalsize\it}} 
\renewcommand\paragraph{\@startsection{paragraph}{4}{\z@}% 
                                   {-1.75ex\@plus -1ex \@minus -.2ex}% 
                                   {1ex \@plus .2ex}% 
                                   {\normalfont\normalsize\bf}} 
\renewcommand\subparagraph{\@startsection{subparagraph}{5}{\z@}% 
                                   {-1.25ex\@plus -0ex \@minus -.2ex}% 
                                   {-2ex \@plus .2ex}% 
                                   {\normalfont\normalsize\it}}

\numberwithin{equation}{section}

\long\def\@makecaption#1#2{%
  \vskip\abovecaptionskip
  \sbox\@tempboxa{{\bf #1:} #2}%
  \ifdim \wd\@tempboxa >\hsize
    {\small\bf #1:} {\small #2}\par
  \else
    \global \@minipagefalse
    \hb@xt@\hsize{\hfil\box\@tempboxa\hfil}%
  \fi
  \vskip\belowcaptionskip}
\renewcommand*\l@section[2]{%
  \ifnum \c@tocdepth >\z@
    \addpenalty\@secpenalty
    \addvspace{.5em \@plus\p@}%
    \setlength\@tempdima{1.5em}%
    \begingroup
      \parindent \z@ \rightskip \@pnumwidth
      \parfillskip -\@pnumwidth
      \leavevmode \bfseries
      \advance\leftskip\@tempdima
      \hskip -\leftskip
      #1\nobreak\hfil \nobreak\hb@xt@\@pnumwidth{\hss #2}\par
    \endgroup
  \fi}
\renewcommand*\l@subsection{\addvspace{.0em \@plus\p@}\@dottedtocline{2}{1.5em}{2.3em}}
\renewcommand*\l@subsubsection{\addvspace{-.2em \@plus\p@}\@dottedtocline{3}{3.8em}{3.2em}}

\def\hepth#1{\href{http://xxx.arxiv.org/abs/hep-th/#1}{{arXiv:hep-th/#1}}}

\def\mathag#1{\href{http://xxx.arxiv.org/abs/math.AG/#1}{{arXiv:math.ag/#1}}}

\def\arxiv#1#2{\href{http://xxx.arxiv.org/abs/#1}{{arXiv:#1 [#2]}}}

\definecolor{refcol}{rgb}{0.2,0.2,0.8}
\definecolor{eqcol}{rgb}{.6,0,0}
\definecolor{purple}{cmyk}{0,1,0,0}

\gdef\@citecolor{refcol}
\gdef\@linkcolor{eqcol}
\def\colorlinkspurple{\gdef\@urlcolor{purple}}
\def\colorlinksblue{\gdef\@urlcolor{blue}}
\def\colorlinksred{\gdef\@urlcolor{red}}

\def\ie{{\it i.e.}}

\def\cf{{\it cf.}}

\def\revise#1       {\raisebox{-0em}{\rule{3pt}{1em}}% 
                     \marginpar{\raisebox{.5em}{\vrule width3pt\ 
                     \vrule width0pt height 0pt depth0.5em 
                     \hbox to 0cm{\hspace{0cm}{% 
                     \parbox[t]{4em}{\raggedright\footnotesize{#1}}}\hss}}}}

\def\ii           {{\it i}}

\def\sqr#1#2{{\vcenter{\vbox{\hrule height.#2pt   
 \hbox{\vrule width.#2pt height#1pt \kern#1pt 
 \vrule width.#2pt}\hrule height.#2pt}}}}

\renewcommand{\P}{\mathbb P}
\newcommand{\R}{\mathbb R}
\newcommand{\Z}{\mathbb Z}
\newcommand{\G}{\mathcal G}
\newcommand{\C}{\mathbb C}
\newcommand{\T}{\mathbb T}

\newcommand{\Ical}{\mathcal I}

\newcommand{\Fcal}{\mathcal F}

\newcommand{\Ocal}{\mathcal O}

\newcommand{\Lcal}{\mathcal L}
\newcommand{\Ccal}{\mathcal C}

\newcommand{\Ncal}{\mathcal N}

\newcommand{\Gcal}{\mathcal G}
\newcommand{\Mcal}{\mathcal M}
\newcommand{\Ecal}{\mathcal E}

\newcommand{\qfrac}{\mathfrak q}

\newcommand{\ep}{\epsilon}
\renewcommand{\t}{\tilde}

\newcommand{\beq}{\begin{equation}}
\newcommand{\eq}{\end{equation}}
\newcommand{\req}[1]{(\ref{#1})}

\catcode`\@=12 

\begin{document} 

%%% 
%%%%%% text starts here 
%%%%%%%%% 

\title{Holomorphic Anomaly in Gauge Theory on ALE space}

\pubnum{
UCB-PTH-11/10
}
\date{December 2011}

\author{
Daniel Krefl and Sheng-Yu Darren Shih  \\[0.2cm]
\it  Center for Theoretical Physics, University of California, Berkeley, USA\\
}

\Abstract{
We consider four-dimensional $\Omega$-deformed $\Ncal=2$ supersymmetric $SU(2)$ gauge theory on $A_1$ space and its lift to five dimensions. We find that the partition functions can be reproduced via special geometry and the holomorphic anomaly equation. Schwinger type integral expressions for the boundary conditions at the monopole/dyon point in moduli space are inferred. The interpretation of the five-dimensional partition function as the partition function of a refined topological string on $A_1\times$(local $\P^1\times\P^1$) is suggested.
}

\makepapertitle

\body

\version\versionno

\vskip 1em

%\tableofcontents

%\newpage

\section{Introduction}
Recently, there has been a renewed interest in gauge theory on Asymptotically Locally Euclidean (ALE) space. This has been mainly triggered by the findings of \cite{BF11a} (see also \cite{BMT11a,BMT11b,W11,I11}), hinting towards an extension of the proposed correspondence between instanton partition functions of $\Ncal=2$ supersymmetric gauge theories in four dimensions and conformal blocks of Liouville theory \cite{AGT09}, to a correspondence between instanton partition functions of gauge theories on ALE space and conformal blocks in super Liouville theory.  

The instanton partition function of four-dimensional $\Ncal=2$ $U(N)$ gauge theory with and without matter can be efficiently calculated using localization on the moduli space of instantons by making use of the so-called $\Omega$-deformation acting on the space-time $\R^4\cong\C^2$ via the rotation
\beq\eqlabel{OmegaDef}
\Omega:\,(z_1,z_2)\rightarrow (e^{\ii \ep_1}z_1,e^{\ii \ep_2}z_2)\,,
\eq
as put forward in \cite{N02} (extending the earlier works \cite{MNS97,LNS98,MNS98}). Subsequently, this localization technique has been applied to gauge instantons on ALE space \cite{FMP04}. Their approach builds on the construction of self-dual gauge connections one ALE space of \cite{KN90}, utilizing that ALE spaces can be obtained from the minimal resolution of orbifolds of type $\C^2/\Gamma$ with $\Gamma$ a discrete Kleinian subgroup of $SU(2)$. Due to the topological nature of the quantities under consideration it is sufficient to stick to the orbifolds, though one may also directly consider the resolved geometries ({\it c.f.}, \cite{BMT11b}).

As shown in \cite{KW10a,KW10b}, $\Omega$-deformed gauge theory partition functions on $\C^2$  can be reproduced under the reparameterization
\beq\eqlabel{epdef}
\ep_1=\sqrt{\beta}\,\lambda\,,\,\,\,\,\ep_2=-\frac{1}{\sqrt{\beta}}\,\lambda\,,
\eq
in the limit $\lambda\rightarrow 0$ with $\beta$ fixed via special geometry and the holomorphic anomaly equation of \cite{BCOV1,BCOV2}. The essential effect of the $\Omega$-deformation being the change of boundary conditions coming from the dyon/monopole point in moduli space. The leading order in $\lambda$ of the free energy, determined via special geometry, is just the prepotential of the gauge theory. Essentially, this is the celebrated Seiberg-Witten solution of $\Ncal=2$ gauge theory \cite{SW94a,SW94b}.  Higher powers in $\lambda$, recursively determined via the holomorphic anomaly equation, supposedly correspond to a 1-parameter family of gravitational corrections.

One may ask if the ``B-model" technique of resorting to special geometry and the holomorphic anomaly equation can be applied as well to the instanton partition functions of gauge theory on ALE space. Due to the linkage between modularity and the holomorphic anomaly \cite{ABK06}, it is not obvious that this is the case, since already in the unrefined case the modular properties of these partition functions are (to the best of our knowledge) somewhat unclear, {\it cf.},  \cite{VW94,BFRM96}. 

As one of our main results we will find that the partition functions of pure $SU(2)\subset U(2)$ gauge theory on the simplest ALE space, namely $A_1$, can be reproduced via the standard B-model approach, albeit supplemented with new boundary conditions. One may take this result as evidence that the applicability of the B-model approach is more general than anticipated. In particular, this indicates that the partition functions of the theories on ALE space still possess modular properties. We will also find that the gauge theory on $A_1$ exhibits a new feature. Namely, the symmetry between the monopole and dyon point may be broken under coupling to gravity. 

In the realm of geometric engineering, the partition function of the undeformed $SU(2)$ gauge theory on $\C^2$ arises as the low energy effective limit of the A-model topological string partition function on local $\P^1\times\P^1$ \cite{KKV96}. For the string theory interpretation of the $\Omega$-deformed gauge theory partition function one usually invokes the M-theory lift. While in the undeformed case the partition function counts BPS states of spinning M2-branes with respect to their left spin under the $SU(2)_L\times SU(2)_R$ rotation group, the $\Omega$-deformation corresponds to a refinement in the sense of counting with respect to left and right spin \cite{HIV03,IKV07}. In particular, the so defined refined topological string partition function equals for geometric engineering geometries the partition function of the corresponding $\Omega$-deformed gauge theory in five dimensions compactified on a circle \cite{HIV03,IKV07}. 

Not very surprisingly, the B-model approach to the $\Omega$-deformation can be applied to the refined topological string partition function, with similar boundary conditions as for the $\Omega$-deformed gauge theory. In particular, the boundary conditions are determined by the $c=1$ string at $R=\beta$ \cite{KW10a}. This is perhaps the strongest hint so far available in favor of the existence of a worldsheet formulation of refinement (for some proposals, see \cite{AHNT10,NO11}). 

Since it is natural to lift the four-dimensional gauge theory partition function on ALE space to five dimensions, in similar fashion as the gauge theory on $\C^2$, it is tempting to speculate about a (refined) topological string interpretation with ALE space-time. In this note we will find some hints towards this direction.  Namely, we will find that the natural five-dimensional lift of the $\Omega$-deformed gauge theory on $A_1$ can be reproduced via the B-model techniques applied to the mirror geometry of local $\P^1\times \P^1$. This suggests that it makes sense to take the five-dimensional gauge theory as a definition of a (refined) topological string on $A_1\times$(local $\P^1\times\P^1$).

The outline is as follows. In the next section we will briefly recall some basics about instanton counting via localization in four dimensions and its generalization to ALE space. The partition function obtained in this manner will be the benchmark to fix the holomorphic ambiguities in the B-model approach, which we will follow in section \ref{BSU2} for $SU(2)\subset U(2)$ gauge theory on $A_1$. The lift of the instanton calculus via localization to five dimensions will be discussed in section \ref{U25DB}. In particular, we will present the natural five dimensional partition functions in the $U(1)$ case, supposedly corresponding to the partition functions of a refined topological string on the resolved conifold with $A_1$ space-time, and for $SU(2)$, which is supposed to correspond to a refined topological string on $A_1\times$(local $\P^1\times\P^1$). The partition functions will be confirmed in section \ref{U25DB} via the B-model approach applied to the mirror geometry of local $\P^1\times\P^1$. We present some concluding words in section \ref{conc}.

\section{Instanton counting via localization}
\label{locSec}
In this section we briefly recall the calculation of the partition function of $\Ncal=2$ supersymmetric gauge theory on $\C^2$ via localization and its generalization to $A_1$ space-time, mainly following \cite{N02,BFRM96,BMT11a}, to there we refer for a more detailed treatment. Though the localization calculation in the general case of $U(N)$ with matter is clear (in the orbifold formalism also with general $A_n$ space-time), for brevity we stick to pure $U(2)$ on $A_1$, as this is the theory of main concern in this note. 

\subsection{Generalities}
According to \cite{N02}, the instanton partition function of $U(2)$ gauge theory on $\R^4\simeq\C^2$ can be calculated via localization with respect to the $\T^2_{a_1,a_2}\times\T^2_{\ep_1,\ep_2}$ group action on the (compactified) moduli space $\Mcal_k$ of $k$-instantons of $U(2)$ gauge theory. Here $\T^2_{a_1,a_2}$ stands for the maximal torus of the $U(2)$ gauge group with generators $a_i$ being the Coloumb parameters, while $\T^2_{\ep_1,\ep_2}$ refers to the maximal torus of the $SO(4)$ space-time rotation group with generators $\ep_i$ as in \req{OmegaDef}. Localization reduces the calculation of the partition function to a weighted summation over the fixed points under the group action $\T^2_{a_1,a_2}\times\T^2_{\ep_1,\ep_2}$. Denoting the set of fixed-points in $\Mcal_k$ as $\Sigma_k\subset\Mcal_k$ and the weights as $\omega_k(f;\vec a;\ep_1,\ep_2)$, the partition function reads
\beq\eqlabel{Zloc}
Z^{\C^2}_{inst}(\vec a;\ep_1,\ep_2;\qfrac)=\sum_{k} \sum_{f\in\Sigma_k}\omega_k[f](\vec a;\ep_1,\ep_2) \, \qfrac^k\,,
\eq
Here, $\qfrac$ stands for the instanton counting parameter, which we will later identify with the dynamical scale of the gauge theory as $\qfrac=\Lambda^4$. The set of fixed-points $\Sigma_k$ can be conveniently encoded in pairs of Young diagrams $\vec Y=(Y_1,Y_2)$ with $|Y_1|+ |Y_2|=k$ \cite{N02}. Furthermore, utilizing the parameterization of the fixed-points via Young diagrams, the weights can be expressed simply as \cite{N02,BFMT02,NY03}
\beq\eqlabel{LocWeight}
\omega_k[\vec Y]=\prod_{n,m=1}^2\frac{1}{\prod_{s\in Y_n}E_s(\ep_1+\ep_2-a_{nm};Y_n,Y_m;-\ep_1,-\ep_2)\prod_{t\in Y_m}E_t(a_{nm};Y_n,Y_m;\ep_1,\ep_2)}\,,
\eq
with $a_{nm}:=a_n-a_m$ and where 
$$
E_s(a,Y_n,Y_m;\ep_1,\ep_2)=a-\ep_1 L_{Y_n}(s)+\ep_2 (A_{Y_m}(s)+1)\,.
$$
Here, $L_Y(s)$ and $A_Y(s)$ denote the usual leg-, respectively, arm-length functions for the box $s=(i,j)$ of the partition $Y$. Note that the partition function $Z^{\C^2}_{inst}$ has to be supplemented by hand by a proper perturbative part $Z_{pert}^{\C^2}$, \ie,
$$
Z^{\C^2}(\vec a;\ep_1,\ep_2;\qfrac)=Z_{pert}^{\C^2}(\vec a;\ep_1,\ep_2) \, Z^{\C^2}_{inst}(\vec a;\ep_1,\ep_2;\qfrac)\,.
$$
Explicit closed formulas for $Z^{\C^2}_{pert}$ can be found in \cite{NO03,NY03b}. Note also that later we will mainly restrict to $SU(2)\subset U(2)$ via setting $a=a_1=-a_2$.

\subsection{$A_1$ via orbifold}
\label{A1viaOrb}
As already mentioned in the introduction, $A_1$ space-time can be obtained as the minimal resolution of the orbifold $\C^2/\Z_2$, where the orbifold, whose  action we denote as $\Ical$, acts on $(z_1,z_2)\in\C^2$ as
\beq\eqlabel{OrbIdef}
\Ical: (z_1,z_2)\rightarrow (e^{\pi\ii} z_1,e^{-\pi\ii} z_2)\,.
\eq 
In particular, the fixed points in the instanton moduli space under the action of $\Omega$ are automatically invariant under the orbifold action. Due to the topological nature of the instanton partition function, one can utilize the orbifold projection to infer the instanton partition function on $A_1$ space-time. Explicitly, this means to restrict the summation in \req{Zloc} to fixed points which are invariant under $\Ical$, combined with a projection of the weights. For that, note that $\Ical$ shifts the $\ep_i$ parameters of the $\Omega$ deformation (\cf, \req{OmegaDef}),  as
\beq\eqlabel{epZnshift}
\ep_1\rightarrow \ep_1+\pi\,,\,\,\,\,\, \ep_2\rightarrow \ep_2-\pi\,.\\
\eq
Because of the orbifold singularity, in general the $U(2)$ gauge field can have non-trivial holonomy when circling around the non-shrinkable cycle. The holonomy can be labeled by a pair of charges $(q_1,q_2)$ associated with the two Cartan generators of the $U(2)$.  Therefore, the Coloumb parameters $a_i$ transform under $\Ical$ as
\beq\eqlabel{ColZnshift}
a_i\rightarrow a_i+\pi q_i\,.
\eq
There are four possible holonomy sectors:  $(0,0)$, $(1,1)$, those partition functions can be reproduced from the Neveau-Schwarz sectors in the associated super Liouville CFT \cite{BF11a}, and $(0,1)$, $(1,0)$ which are related to the Rammond sector of the CFT \cite{I11}. All other fields, being covariantly constant with respect to the gauge field, acquire a phase shift  induced from the twisted boundary condition of the gauge field when going around the non-trivial cycle.

As discussed in detail in \cite{FMP04}, the projection on the fixed-points can be implemented by associating to each box $(i,j)$ in a Young diagram $Y_n$ a charge 
$q_n +i-j$ and counting the total number of boxes with zero charge, $k_0$, and of charge one, $k_1$, for each pair of Young diagrams (corresponding to a fixed-point). The topological classification of the instanton solution on $\C^2/\Z_2$ according to its first and second Chern class \cite{KN90} then yields selection rules on the allowed values of $(k_0,k_1)$ for given holonomy. Explicitly, we take as in \cite{I11} for the four different sectors of $U(2)$ on $\C^2/\Z_2$:
$$
(0,0),(0,1),(1,0): k_0-k_1=0\,,\,\,\,\, (1,1): k_0-k_1=1\,.
$$
Note that it may seem that this corresponds only to a small subset of possible instanton solutions. However, we will see in the next section that this subset in fact contains all non-trivial information. 

It remains to project the weights. The projection is simply given by keeping only the factors in \req{LocWeight} which are even under $\Ical$, acting via \req{epZnshift} and \req{ColZnshift} on each factor. Following this prescription, it is straight-forward to calculate the partition function $Z^{(q_1,q_2)}$ for each holonomy sector. 

We reparameterize the partition functions in the variables \req{epdef}, restrict to $SU(2)\subset U(2)$ and expand the partition functions in $\lambda$, keeping $\beta$ fixed. The expansion goes into even powers of $\lambda$ only, \ie,
\beq\eqlabel{Fdef}
\Fcal(a;\ep_1,\ep_2;\qfrac):=\log Z(a;\ep_1,\ep_2;\qfrac)\sim \sum_{n=0}^\infty \Gcal^{(n)}(a;\beta;\qfrac)\,\lambda^{2n-2}\,.
\eq
For the applicability of the usual Seiberg-Witten approach, we would like to have that
\beq\eqlabel{G0F0rel}
\Gcal^{(0)}\sim\Fcal^{(0)}\,,
\eq
(up to some rescaling of $\lambda$) where $\Fcal^{(0)}$ is the ordinary Seiberg-Witten prepotential of $SU(2)$ gauge theory on $\C^2$. In physical language, this means that we would like to see a difference only under coupling to gravity, which is responsible for the higher order terms in $\lambda$ in \req{Fdef}. Extracting explicitly the prepotential $\G^{(0)}$ for each sector $Z^{(q_1,q_2)}$, however, shows that this is generally not the case. Rather, one has to take the combinations of holonomy sectors
\beq\eqlabel{ZpmDef}
\begin{split}
Z^{+}_2&:=Z^{(0,1)}+Z^{(1,0)}\,,\\
Z^{-}_2&:=Z^{(0,0)}+Z^{(1,1)}\,,
\end{split}
\eq
in order that \req{G0F0rel} holds (strictly speaking, taking the combination leading to $Z^+_2$ is not necessary since $Z^{(0,1)}=Z^{(1,0)}$). That is, one should combine sectors with same first Chern class. Note that while $Z_2^+$ obtains contributions only from regular instantons, $Z_2^-$ includes contributions from regular and fractional instantons.

Hence, $SU(2)\subset U(2)$ gauge theory on $A_1$ possesses two distinguished sectors under coupling to gravity, with partition functions as defined via \req{ZpmDef}. (Sometimes we will also refer to $Z^+_2$ as coming from the even sector and $Z^-_2$ as coming from the odd sector.) These are the two combinations of sectors which are well-behaved (in the sense of having modular properties), and on which we therefore focus on in this work.

\subsection{$A_1$ via blowup}
While the orbifold approach has the advantage that it is rather simple, efficient to explicitly compute and straight-forward to generalize, it has one obvious drawback. Namely, it is not immediately clear how to project the perturbative part of the partition function. However, calculating the partition function instead via a more algebraic geometrical approach one can in fact obtain a prediction. The price to pay is that this approach is somewhat more complicated and less efficient to explicitly compute. Therefore, we will mainly use the orbifold approach for explicit calculations and use the algebraic approach only to gain some additional insight. The construction goes (roughly) as follows \cite{BMT11a} (see also \cite{BPT09} and references therein).
  
The compactification of the instanton moduli space of $U(2)$ gauge theory on the minimal resolution $\Ocal(-2)\rightarrow\P^1$ of the $A_1$ singularity can be described in terms of the moduli space $\Mcal(2,k,n)$ of a rank two framed torsion free coherent sheave $\Ecal$ on a stacky compactification of $\Ocal(-2)\rightarrow\P^1$. The moduli space is characterized by the first Chern class $c_1(\Ecal)=k E$ (with $E$ the exceptional divisor resolving the singularity) and the discriminant $\Delta(\Ecal)=c_2(\Ecal)-\frac{1}{4}c_1^2(\Ecal)=n$. It is important to keep in mind that due to the stacky compactification $k$ can be integer and half-integer (\cf, \cite{BPT09}). As usual, the instanton partition function is obtained via localization with respect to a $\T^2_{a_1,a_2}\times\T^2_{\ep_1,\ep_2}$ action on the moduli space, as a weighted sum over the fixed-points. The fixed-points under the torus action are given by (twisted) ideal sheaves characterized by a pair $\vec k=(k_1,k_2)$ with $k=k_1+k_2$ and a pair of Young diagrams $(Y_1,Y_2)$ with $|Y_1|+|Y_2|=n+\frac{1}{2}(k_1-k_2)^2$. Though it is not hard to infer the weights, they are too lengthy to explicitly recall here. The details of the localization calculation are anyway not of our main concern in this note. Therefore, we just state the final result for the instanton partition function, which takes the form \cite{BMT11a}
\beq\eqlabel{ZinstA1}
Z^{A_1}_{inst}=\sum_{k}\qfrac^{|k|^2/2}  Z^{(k)}(\vec a;\ep_1,\ep_2;\qfrac)\,,
\eq
with
\beq\eqlabel{ZinstA1k}
\begin{split}
Z^{(k)}(\vec a;\ep_1;\ep_2;\qfrac)=&\sum_{k_1+k_2=k}\qfrac^{|k_1-k_2|^2/2}\,\prod_{\alpha,\beta=1}^2\ell(\vec a;k_\alpha,k_\beta; \ep_1,\ep_2)\\
&\times Z^{\C^2}_{inst}(\vec a-2\ep_1 \vec k;2\ep_1,\ep_2-\ep_1;\qfrac) Z^{\C^2}_{inst}(\vec a-2\ep_2 \vec k;\ep_2-\ep_1,2\ep_2;\qfrac)\,,
\end{split}
\eq
and 
$$
\ell(\vec a;k_\alpha,k_\beta; \ep_1,\ep_2)=\left\{
\begin{matrix}
\prod_{\sigma_+} \frac{(-1)}{i\ep_1+j\ep_2}&&k_{\alpha\beta}>0\\
\prod_{\sigma_-} \frac{1}{(i+1)\ep_1+(j+1)\ep_2}&&k_{\alpha\beta}<0\\
1&&k_{\alpha\beta}=0
\end{matrix}
\right.\,,
$$
where we defined $k_{\alpha\beta}=k_\alpha-k_\beta$ and $Z^{\C^2}_{inst}$ is as defined via \req{Zloc}. The selection rules on the products are $\sigma_+:=\{i,j\geq 0:i+j\leq 2(k_{\alpha\beta}-1)\wedge i+j-2k_{\alpha\beta}=0\mod 2\}$ and $\sigma_-:=\{i,j\geq 0:i+j\leq -2(k_{\alpha\beta}+1)\wedge i+j-2k_{\alpha\beta}=0\mod 2\}$.

Some remarks are in order. The so defined partition function \req{ZinstA1} is a summation over all topological sectors. In order to compare to the orbifold partition functions of the previous section, one has to identify the corresponding values of $k$. It is not hard to infer that
\beq
Z^{(0)}(\vec a;\ep_1;\ep_2;\qfrac)=Z_2^-\,,\,\,\,\,\, Z^{(1/2)}(\vec a;\ep_1;\ep_2;\qfrac)=\qfrac\, Z_2^+\,,
\eq
with $Z^\pm_2$ as defined in \req{ZpmDef}. This is as expected since fractional $k$ corresponds to non-trivial holonomy (\cf, \cite{BPT09}). Furthermore, one can in fact observe that the sum over the integer and half-integer sectors of \req{ZinstA1k} factorize separately, \ie, 
$$
Z^{A_1}_{inst}=\vartheta_3(0;\qfrac^{1/2})\, Z_2^- +\qfrac^{-1/8}\vartheta_2(0;\qfrac^{1/2}) \,Z_2^+\,,
$$
where $\vartheta_n(z;q)$ denote the standard auxiliary theta functions. A related observation has been made for the $\Ncal=2^*$ theory in \cite{BMT11a}, and one may also obtain the above factorization via taking an appropriate limit thereof. This explains the statement of the previous section that only the topological sectors leading to $Z_2^+$ and $Z_2^-$ yield non-trivial information and are thereof of relevance to us. 

Finally, we also briefly recall from \cite{BMT11a} that the form of the partition function \req{ZinstA1} can be used to predict the perturbative contribution. The main idea behind is that if one requires that \req{ZinstA1k} is expressed in a similar fashion in terms of the full partition function $Z^{\C^2}$ (\ie, that a similar blowup equation as in \cite{NY03} holds), the perturbative contribution is essentially fixed and can be extracted. 

\section{B-model approach to $SU(2)$ gauge theory in 4d}
\label{BSU2}
In this section we are going to reproduce the partition functions of the gauge theory on $A_1$ space of the previous section using special geometry and the holomorphic anomaly equation. In particular, we will infer the behavior at the dyon/monopole point in moduli space. Also, this approach definitely fixes the perturbative contribution, for which we will find closed and simple expressions. 

\subsection{Generalities}

We consider $\Ncal=2$ $SU(2)$ gauge theory in four dimensions and denote by $u$ a global coordinate on the quantum moduli space of vacua $\Mcal$, which can be identified with the base space of a family of complex curves $\Ccal_u$. For instance the hyperelliptic curve, 
$$
\Ccal_u:~y^2=(x^2-\Lambda^4)(x-u)\,,
$$
where $\Lambda$ denotes the dynamical scale of the gauge theory. The quantum moduli space has three special points. Namely, the weak coupling region $u=\infty$, where the gauge boson (vector-multiplet) become massless, the strongly coupled monopole point $u=\Lambda^2$ with a massless monopole (hyper-multiplet) and the strongly coupled dyon point $u=-\Lambda^2$ with a massless dyon (hyper-multiplet). Note that the monopole and dyon points are related by a $\Z_2$ symmetry. For latter reference, we parameterize the strongly coupled singular points via the ``discriminant" $\Delta=0$ with
\beq\eqlabel{Dis}
\Delta=\Delta_+ \Delta_-\,,
\eq
and 
\beq\eqlabel{Dispm}
\Delta_\pm=u\pm\Lambda^2\,.
\eq
The family of curves is equipped with a meromorphic one-form $\lambda_{SW}$, such that 
$$
a=\oint_A \lambda_{SW}\,,\,\,\,\,\, a_D=\oint_{A_D}\lambda_{SW}\,,
$$
and
$$
a_D=\frac{\partial\Fcal^{(0)}}{\partial a}\,,
$$
for appropriately chosen 1-cycles $A$ and $A_D$, and under elimination of $u$, via the so-called mirror map $u(a)$ obtainable by inverting the period $a(u)$.
 In order to obtain the periods $a$ and $a_D$ in different corners of moduli space, it is convenient to resort to the Picard-Fuchs equation, 
$$
\Lcal\, \omega(u)=0\,,
$$
satisfied by all the periods of $\lambda_{SW}$. Taking $a$ as local (flat) coordinate around $u\rightarrow\infty$, the differential operator $\Lcal$ takes the form
$$
\Lcal=\partial_a\frac{1}{C_{aaa}}\partial_a^2\,,
$$
where $C_{aaa}:=\partial_{a}^3\Fcal^{(0)}$ is referred to as Yukawa coupling. 

We now apply the holomorphic anomaly equations of \cite{BCOV1,BCOV2}. That is, we consider the amplitudes $\Gcal^{(2n-2)}(a;\beta;\qfrac)$ (defined in \req{Fdef}) to be given by the holomorphic limit $\bar a\rightarrow\infty$ of non-holomorphic, but globally defined objects $\Gcal^{(n)}(u,\bar u)$ over $\Mcal$. 

For $g>1$, the $\Gcal^{(g)}(u,\bar u)$ satisfy the holomorphic anomaly equation
\beq\eqlabel{holaneq}
\bar\partial_{\bar u}\Gcal^{(g)}=\frac{1}{2} \sum_{r=1}^{g-1}\bar C_{\bar u}^{uu}\Gcal_{u}^{(g-r)}\Gcal_{u}^{(r)}+\frac{1}{2}\bar C_{\bar u}^{uu}\Gcal_{uu}^{(g-1)}\,,
\eq
where $\Gcal^{(g)}_{uu}=D_u\Gcal^{(g)}_u=D_u^2\Gcal^{(g)}$ and $D_u$ is the covariant derivative over $\Mcal$. In particular, the connection takes the form
$$
\lim_{\bar a\rightarrow \infty}\Gamma^u_{uu}=\partial_u\log\frac{\partial a(u)}{\partial u}\,.
$$
As usual, inices are raised and lowered using the Weil-Petersson metric. The one-loop amplitude satisfies
\beq\eqlabel{1holan}
\bar \partial_{\bar u}\partial_u\Gcal^{(1)}=\frac{1}{2}\bar C^{uu}_{\bar u} C_{uuu}\,.
\eq
There are several techniques on the market to solve the holomorphic anomaly equation \req{holaneq}. Hence, we will not dwell here into the details on how to actually solve \req{holaneq}. For that, we refer to the (extensive) literature on this topic. 

Clearly, $\req{holaneq}$ and $\req{1holan}$ only capture the non-holomorphic part of the amplitude and hence have to be supplemented by an appropriately chosen holomorphic function. The determination of the holomorphic function is the main difficulty in the holomorphic anomaly approach. Fortunately, the holomorphic ambiguites can be fixed by taking boundary conditions from other points in moduli space into account \cite{HK06,HKQ06}. In particular, for $SU(2)$ gauge theory in the $\Omega$ background, expansion of the amplitudes at the monopole or dyon point is sufficient, since the amplitudes feature at this point in moduli space a gap with a specific singular leading term, yielding enough boundary conditions. In detail, the leading terms are determined by the free energy of the $c=1$ string at $R=\beta$ \cite{KW10a}.

Let us pause for a moment to comment on a related approach in the literature. For the same purpose of reproducing refined partition functions via holomorphic anomaly equations the authors of \cite{HK10} promote a ``generalized" version of the holomorphic anomaly equation. The reason behind lies in their two parameter expansion of the free energy $\Fcal(a;\ep_1,\ep_2;\qfrac)$ (as defined in \req{Fdef}) , \ie,  
$$
\Fcal(a;\ep_1,\ep_2;\qfrac)=\sum_{n_1,n_2=0}^\infty\Fcal^{(n_1,n_2)}(a;\qfrac)(\ep_1\ep_2)^{n_1-1}(\ep_1+\ep_2)^{2n_2}\,.
$$
However, one can infer that the relation to our 1-parameter expansion discussed above is just a resummation,
$$
\Gcal^{(g)}(a;\beta;\qfrac)=\sum_{(n_1,n_2)\subset\sigma(g,_2)}\left(\sqrt{\beta}-\frac{1}{\sqrt{\beta}}\right)^{2n_2}\Fcal^{(n_1,n_2)}(a;\qfrac)\,,
$$ 
where $\sigma(g,2)$ denotes the set of integer (including zero) partitions of $g$ of length $2$. In particular, their ``generalized" holomorphic anomaly equation follows from the above relation. Hence, the refinement is generally captured by the ordinary holomorphic anomaly equation equipped with new boundary conditions, as put forward in \cite{KW10a}. For completeness, we also mention that one can also consider the extended holomorphic anomaly equation of \cite{W07} in this context in order to capture certain shift degree of freedoms, as also discussed in \cite{KW10a,KW10b}.

\subsection{$Z_2^+$}
Let us start the discussion of the gauge theory on $A_1$ space with the sector with partition function $Z^+_2$. As already stated in section \ref{A1viaOrb}, via explicit expansion of the instanton partition function as in \req{Fdef}, one infers that 
$$
\Gcal^{(0)}=\frac{1}{2}\Fcal^{(0)}\,,
$$
and hence we can work with the usual special geometry of $SU(2)$ gauge theory (under a rescaling of $\lambda$). Furthermore, we observe that the 1-loop sector is reproduced via the usual solution of the 1-loop holomorphic anomaly equation \req{1holan}, given by
\beq\eqlabel{Z011loop}
\Gcal^{(1)}=\frac{1}{2}\log\left(\partial_u a(u)\right)+a^{(1)}_\Gcal(u;\beta)\,,
\eq
with holomorphic ambiguity fixed to
$$
a^{(1)}_\Gcal(u;\beta)=\frac{1}{48}\left(\beta+\frac{1}{\beta}\right)\log\Delta+\frac{1}{8}\log\Delta\,.
$$
It is interesting to note that this holomorphic ambiguity seems also to be related to the 1-loop partition function of a specific two dimensional gravity model, \cf, \cite{BK90}, similar as is the case for the ordinary $\Omega$-deformed 1-loop amplitude. 

For higher genus, we parameterize the holomorphic ambiguities as usual via the functions
$$
a^{(g>1)}_\Gcal(u;\beta)=u^{3-g}\sum_{n=1}^{2g-2} a_n^{(g)}(\beta) \frac{\Lambda^{4n-4}}{\Delta^n}\,,
$$
with constants $a_n(\beta)$ to be determined. 

Via explicit expansion of the higher $\Gcal^{(g)}$ at the weak coupling and monopole/dyon points, and comparing with the expected results from the localization calculation outlined in section \ref{A1viaOrb}, we are able to fix the holomorphic ambiguities, that is, the $a^{(g)}_n(\beta)$. In particular, we observe that the genus $g$ amplitudes expanded at the monopole/dyon points have the structure  
\beq
\Gcal^{(1)}(a_D;\beta)=\Psi^{(1)}_2(\beta)\,\log(a_D)+\dots\,,\,\,\,\,\,\Gcal^{(g>1)}(a_D;\beta)=\Psi^{(g)}_2(\beta)\, a_D^{{2-2g}}+\Ocal(a_D^{0})\,,
\eq
\ie, feature a gap and a distinguished leading singular term paramterized as $\Psi_2^{(g)}$. We find for the first few of the coefficients
\beq\eqlabel{Psis}
\begin{split}
\Psi_2^{(1)}(\beta)&=\frac{1}{48}\left(\beta+\frac{1}{\beta}\right)+\frac{1}{8}\,,\\
\Psi_2^{(2)}(\beta)&=-\frac{7+180\beta+10\beta^2+180\beta^3+7\beta^4}{11520\beta^2}\,,\\
\Psi_2^{(3)}(\beta)&=\frac{31+3150\beta+49\beta^2+3780\beta^3+49\beta^4+3150\beta^5+31\beta^6}{322560\beta^3}\,,\\
&~~\vdots
\end{split}
\eq

We can also extract the perturbative contribution to the partition function, which we parameterize as
$$
\log Z_{2}^{pert}\sim 2\,\Phi_2^{(1)}(\beta)\log a+\sum_{n >1}2\, \Phi_2^{(n)}(\beta)\, \frac{\lambda^{2n-2}}{(2a)^{2n-2}}\,.
$$
We find
\beq\eqlabel{Phis}
\begin{split}
\Phi_2^{(1)}(\beta)&=-\frac{1}{24}\left(\beta+\frac{1}{\beta}\right)\,,\\
\Phi_2^{(2)}(\beta)&=\frac{1+40\beta^2+\beta^4}{1440\beta^2}\,,\\
\Phi_2^{(3)}(\beta)&=-\frac{1+154\beta^2+154\beta^4+\beta^6}{10080\beta^3}\,,\\
\Phi_2^{(4)}(\beta)&=\frac{3+1880\beta^2+1568\beta^4+1880\beta^6+3\beta^8}{60480\beta^4}\,,\\
&~~\vdots
\end{split}
\eq

Note that we can reproduce both the $\Psi_2^{(g)}(\beta)$ and $\Phi_2^{(g)}(\beta)$ via the Schwinger type integrals \footnote{The reader should not be confused about $\sinh$ versus $\sin$, respectively $\cosh$ versus $\cos$ in comparision to other Schwinger type integrals in the literature. What occurs depends on the definition/convention used.} 
\beq\eqlabel{SchwingerInt}
\begin{split}
\frac{1}{2}\int\frac{ds}{s}e^{-s\mu}\frac{\cosh\left(\frac{s(\ep_1+\ep_2)}{2}\right)}{\sinh(s\ep_1)\sinh(s\ep_2)} &\sim\dots+\sum_{n>0}\Psi_2^{(n)}(\beta)
\,\frac{\lambda^{2n-2}}{\mu^{2n-2}} \,,\\
\frac{1}{2}\int\frac{ds}{s}e^{-s\mu}\frac{e^{\frac{s}{2}(\ep_2-\ep_1)}\cosh\left(\frac{s(\ep_1+\ep_2)}{2}\right)}{\sinh(s\ep_1)\sinh(s\ep_2)} &\sim\dots+\sum_{n>2}\Phi_2^{(n/2)}(\beta)\,
\frac{\lambda^{n-2}}{\mu^{n-2}} \,,
\end{split}
\eq
where the relation between the first and second integral is just a shift of $\mu$, 
\beq\eqlabel{shift1}
\mu\rightarrow\mu +\frac{1}{2}(\ep_1-\ep_2)\,.
\eq
One should note that the Schwinger integral for $\Phi_2$ presented in equation \req{SchwingerInt} has an additional sector in odd powers of $\lambda$ and is further not symmetric in the exchange of the $\ep_i$. Both problems can be evaded if one considers instead the integral 
\beq\eqlabel{SchwingerIntvec}
\frac{1}{4}\int\frac{ds}{s}e^{-s\mu}\frac{\cosh\left(\frac{s(\ep_1+\ep_2)}{2}\right)\cosh\left(\frac{s(\ep_1-\ep_2)}{2}\right)}{\sinh(s\ep_1)\sinh(s\ep_2)}\,.
\eq
The odd sector disappears while the even sector is unchanged. Hence,  one should interpret the first integral in \req{SchwingerInt} and the integral in \req{SchwingerIntvec} as due to integrating out a massive hyper-, respectively, vector-multiplet in the $\Omega$-deformed $A_1$ background. In particular, one can see the vector-multiplet contribution \req{SchwingerIntvec} as the combination of two hyper-multiplets shifted with the shift \req{shift1} with opposite signs, as it should be since we specialized to $SU(2)\subset U(2)$. Note that in contrast to gauge theory in ordinary space-time  (\cf, \cite{KW10b}), the hyper- and vector-multiplets give already for $\ep_1=-\ep_2$ different contributions.

Since we have at genus $g$ exactly $2g-2$ unknowns $a^{(g)}_n$, the gap provides $2g-1$ boundary condition and we have a closed expression for the $\Psi_2^{(2)}$, we can (at least theoretically) solve for the partition function to any desired order in the B-model formalism. 

Let us also comment on the Nekrasov-Shatashvilli limit (for short NS limit) \cite{NS09} of the gauge theory partition function in the $A_1$ case. Since the partition function is entirely determined by the Schwinger integrals above, it is sufficient to consider the NS limit thereof. We have
$$
\lim_{\ep_2\rightarrow 0}\ep_2\int\frac{ds}{s}e^{-s\mu}\frac{\cosh\left(\frac{s(\ep_1+\ep_2)}{2}\right)}{\sinh(s\ep_1)\sinh(s\ep_2)}=\frac{1}{2}\int\frac{ds}{s^2}\frac{e^{-\mu s}}{\sinh\left(\frac{s\ep_1}{2}\right)}\,.
$$
Comparision with \cite{ACDKV11} shows that in the NS limit the boundary conditions, hence also the full partition function, of the $A_1$ case becomes identical to the ordinary case (up to an overall factor of $1/2$). In particular, this implies that the quantum integrable system one can associate in the NS limit with the $\Omega$-deformed gauge theory, following \cite{NS09}, is as for the gauge theory on $\C^2$. 

\subsection{$Z^-_2$}
Let us now discuss the other sector. Similar as before, we have that
$$
\Gcal^{(0)}=\frac{1}{2} \Fcal^{(0)}\,,
$$
and the usual special geometry applies. However, in contrast to above, we find that the 1-loop amplitude is reproduced via the ambiguity in \req{Z011loop} fixed to 
\beq\eqlabel{U20011a1}
a_\Gcal^{(1)}=\frac{1}{48}\left(\beta+\frac{1}{\beta}\right)\log\Delta+\frac{1}{8}\log\Delta_+-\frac{1}{8}\log\Delta_-\,.
\eq
In particular, this shows that the $\Z_2$ symmetry between the monopole and dyon point in moduli space at $u=\pm\Lambda^2$ is broken at 1-loop. This is similar as in the shifted two massless flavor case considered in \cite{KW10a}. Correspondingly, in order to obtain sufficient boundary conditions we have to expand the higher genus amplitudes separately around both points in moduli space. We take as Ansatz for the holomorphic amiguities 
$$
a_\Gcal^{(g)}(u;\beta)=u^{3-g}\sum_{n=1}^{2g-2}\frac{1}{\Delta^n}\left( a^{(g)}_n(\beta) \Lambda^{4n-4}+\t a^{(g)}_n(\beta) u^{2n+1}\Lambda^{2}\right)\,,
$$
where we assumed twice as much unknowns, $a_n^{(g)}$ and $\t a_n^{(g)}$, as we had before because of the breaking of the $\Z_2$ symmetry. Fixing the unknown coefficients via comparision with the localization results and analytic continuation of the amplitudes to the monopole and dyon point shows that the amplitudes still feature a gap structure, \ie, 
\beq
\Gcal^{(1)}_\pm (a_D;\beta)=\Psi^{(1)}_2(\pm\beta)\,\log(a_D)+\dots\,,\,\,\,\,\,\Gcal_\pm^{(g>1)}(a_D;\beta)=\Psi^{(g)}_2(\pm \beta)\, a_D^{{2-2g}}+\Ocal(a_D^{0})\,,
\eq
where the $\pm$ distinguishes between expansion at the monopole, respectively, dyon point in moduli space. We observe that the coefficients $\Psi_2$ are as in \req{Psis}. The sole difference between the expansions at the two points is a flip of sign of $\beta$. 

We can also read of the perturbative contribution. The first few terms read
\beq
\begin{split}
\t\Phi^{(1)}_2(\beta)&=-\frac{1}{24}\left(\beta+\frac{1}{\beta}\right)+\frac{1}{4}\,,\\
\t\Phi^{(2)}_2(\beta)&=\frac{1-50\beta^2+\beta^4}{1440\beta^2}\,,\\
\t\Phi^{(3)}_2(\beta)&=-\frac{1-161\beta^2-161\beta^4+\beta^6}{10080\beta^3}\,,\\
&~~\vdots
\end{split}
\eq 
It is instructive to write down the corresponding Schwinger integrals, which are closely related to the ones given in \req{SchwingerInt}. We have
\beq\eqlabel{SchwingerInt2}
\begin{split}
\frac{1}{2}\int \frac{ds}{s}e^{-\mu s}\frac{\cosh\left(\frac{s(\ep_1\pm\ep_2)}{2}\right)}{\sinh(s\ep_1)\sinh(s\ep_2)}&\sim\dots+\sum_{n>0}\Psi_2^{(n)}(\pm\beta)
\,\frac{\lambda^{2n-2}}{\mu^{2n-2}} \,,\\
\frac{1}{2}\int \frac{ds}{s}e^{-\mu s}\frac{e^{\frac{1}{2}s(\ep_1-\ep_2)}\cosh\left(\frac{s(\ep_1+\ep_2)}{2}\right)}{\sinh(s\ep_1)\sinh(s\ep_2)}&\sim\dots+\sum_{n>0}\t\Phi_2^{(n/2)}(\beta)\,
\frac{\lambda^n}{\mu^n} \,.
\end{split}
\eq
Note that the second integral in \req{SchwingerInt2} is related to the first integral with plus sign via a shift 
\beq\eqlabel{shift2}
\mu\rightarrow \mu-\frac{1}{2}(\ep_1+\ep_2)\,.
\eq
As in the previous subsection, one should get rid off the odd sector in the integral for $\t\Phi_2$ via considering instead the integral
$$
\frac{1}{4}\int \frac{ds}{s}e^{-\mu s}\frac{\cosh^2\left(\frac{s(\ep_1+\ep_2)}{2}\right)}{\sinh(s\ep_1)\sinh(s\ep_2)}\,.
$$
Again, this can be seen as the combination of two hyper-multiplets oppositely shifted via \req{shift2}.

Since we have at genus $g$ in total $4g-4$ unknowns $a^{(g)}_n$ and $\t a^{(g)}_n$, the two gaps provide $4g-2$ boundary conditions, and the Schwinger integrals in \req{SchwingerInt2} provide $3$ additional boundary conditions, we have again enough boundary information available to calculate the partition function in the B-model formalism to any desired order. 

We also remark that trivially the $\Psi^{(n)}_2(\pm\beta)$ have the same $\ep_2\rightarrow 0$ limit for both signs. Hence, the statement of the previous section regarding the NS limit still applies. 

Finally, let us note that due to the identity
\beq\eqlabel{cossinid}
\frac{\cosh\left(\frac{s(\ep_1+\ep_2)}{2}\right)+\cosh\left(\frac{s(\ep_1-\ep_2)}{2}\right)}{\sinh(s\ep_1)\sinh(s\ep_2)}=\frac{1}{2\sinh(\frac{s\ep_1}{2})\sinh(\frac{s\ep_2}{2})}\,,
\eq
we have that
\beq\eqlabel{PsiRelation}
\Psi_1^{(g)}(\beta)=\Psi^{(g)}_2(\beta)+\Psi_2^{(g)}(-\beta)\,,
\eq
where $\Psi_1^{(g)}(\beta)$ denote the usual boundary conditions of $SU(2)$ gauge theory on $\Omega$-deformed $\C^2$ space-time at the monopole/dyon point \cite{KW10a}.

The relation \req{PsiRelation} neatly illustrates the essential point of the B-model discussion performed in this section. Namely, the boundary conditions from the dyon/monopole points are projected to either $\Psi_2^{(g)}(+\beta)$ or $\Psi_2^{(g)}(-\beta)$, separately for each point. This freedom incorporates the two different sectors of the gauge theory on $A_1$ space-time outlined in section \ref{A1viaOrb} into the B-model formalism.

\section{Instanton partition functions in 5d}
In this section we are going to consider the natural lift of the localization calculation for four-dimensional gauge theory of section \ref{locSec} to five dimensions. In particular, we take the so-obtained partition functions as definition of a refined topological string with $A_1$ space-time.
\subsection{Generalities}
The partition function of five-dimensional supersymmetric gauge theory with eight supercharges compactified on a circle of radius $R_c$ can be obtained via a certain deformation of the instanton calculation for the same gauge theory in four dimensions \cite{N02,NO03,NY05}. In detail, the deformation is given by a simple change of weights
\beq\eqlabel{4dto5d}
E_s(a;Y_n,Y_m;\ep_1,\ep_2) \rightarrow 1-\exp \left(R_c\, E_s(a;Y_n,Y_m;\ep_1,\ep_2) \right)\,,
\eq
in the localization formula \req{LocWeight}. We can absorb $R_c$ in a simultaneous rescaling of the Coloumb-parameters $a_i$ and the parameters $\ep_i$. Therefore we just set $R_c$ to one in \req{4dto5d}. 

By definition, the partition function of the five-dimensional gauge theory can be identified with the partition function of a refined topological string on a geometric engineering geometry which yields in the effective field-theory limit the four-dimensional gauge theory partition function \cite{HIV03,IKV07}. 

It seems natural to conjecture that the ``lift"  \req{4dto5d} applied to the gauge theory on ALE space also corresponds to the partition function of a five dimensional gauge theory, presumably on ALE$\times S^1$ (at least locally), which, one may take as the definition of a refined topological string with ALE space-time. 

\subsection{$U(1)$}
Let us consider first the $U(1)$ gauge theory in five dimensions, those partition function defines the refined topological string on the conifold, that is, $\Ocal(-1)\oplus\Ocal(-1)\rightarrow\P^1$ \cite{HIV03}. Since we are considering $U(1)$ gauge theory, we only have to sum over a single partition $R$. Before orbifolding, one has \cite{NY03,NY05}
$$
Z_{U(1)}=\sum_R \qfrac^{|R|} C_{R} C_{R^t}\,,
$$
with
\beq\eqlabel{Cdef}
C_{R}:=\prod_{(i,j)\in R}\left(1-e^{\ep_1L_R(i,j)-\ep_2(A_R(i,j)+1)}\right)^{-1}\,,
\eq
where as before $L_R(i,j)$ denote the leg-length, $A_R(i,j)$ the arm-length of the box $(i,j)\in R$ and $\qfrac$ the instanton counting parameter, which is identified with the K\"ahler parameter $t$ of the geometry as 
$$
\qfrac =e^{\frac{1}{2}(\ep_1+\ep_2)}Q\,,
$$
where we also defined $Q:=e^{-t}$. Note that we performed an additional shift of $t$ in order to obtain an expansion of the corresponding free energy into even powers of $\lambda$ only. In particular, under this identification one can show that 
\beq\eqlabel{ZconiExp}
Z=\exp\left(\sum_{k=1}^\infty \frac{Q^k}{4k\sinh(\frac{k\ep_1}{2})\sinh(\frac{k\ep_2}{2})}\right)=\prod_{i,j}^\infty(1-Q e^{\ep_1(i-1/2)-\ep_2(j-1/2)})\,,
\eq
and this is the refined partition function we expect from the M-theory spin state counting point of view, following \cite{HIV03}. Note also that the partition function of $\Ocal(-2)\oplus\Ocal(0)\rightarrow \P^1$ differs from this only by $Z\rightarrow Z^{-1}$. 

Let us now consider the orbifold projection. Following the implementation of the projection into the localization scheme, as described in section \ref{A1viaOrb}, we infer that we obtain as projected partition function $Z_{U(1)}^+$,
\beq\eqlabel{ConiZviaVertex}
Z_{U(1)}^+=\sum_{R_+}  \qfrac^{|R_+|/2} \tilde C_{R_+} \tilde C_{R_+^t}\,,
\eq
where $R_+$ denotes the projected set of partitions with an even number of boxes, and  $\tilde C_{R}$ is the projection of \req{Cdef}, that is,
only boxes in $R$ are taken to contribute for which 
$$
L_R(i,j)+A_R(i,j)+1=0 \mod 2\,,
$$
holds.

In order to find a similar expression as \req{ZconiExp} for the orbifold, a natural guess for an Ansatz would be to directly apply the orbifold projection to the infinite product occuring in \req{ZconiExp}. Let us assume that the K\"ahler modulus $t$ is charged under the orbifold action $\Ical$ given in \req{OrbIdef} with charge $q^c \pi$. The product is invariant under $\Ical$ if
$$
(i+j-1)+q^c=0\mod 2\,,
$$
holds. For $q^c$ even, the projected partition function is given by
$$
Z_{U(1)}^+=\prod_{i_1,j_2\,{\rm even}}^\infty\prod_{j_1,i_2\,{\rm odd}}^\infty(1-Q e^{\ep_1(i_1-1/2)-\ep_2(j_1-1/2)})(1-Q e^{\ep_1(i_2-1/2)-\ep_2(j_2-1/2)})\,,
$$
while for $q^c$ odd by
$$
\t Z_{U(1)}^-=\prod_{i_1,j_1\,{\rm even}}^\infty\prod_{j_2,i_2\,{\rm odd}}^\infty(1-Q e^{\ep_1(i_1-1/2)-\ep_2(j_1-1/2)})(1-Q e^{\ep_1(i_2-1/2)-\ep_2(j_2-1/2)})\,.
$$
We can rewrite the first partition function as 
\beq\eqlabel{ConiZviaProd}
Z^+_{U(1)}=\exp\left(\sum_{k=1}^\infty\frac{e^{\frac{1}{2}k(\ep_1+\ep_2)}(1+e^{k(\ep_1+\ep_2)})}{k(e^{2k\ep_1}-1)(e^{2k\ep_2}-1)}Q^k\right)=\exp\left(\sum_{k=1}^\infty\frac{\cosh\left(\frac{k(\ep_1+\ep_2)}{2}\right)}{2k\sinh(k\ep_1)\sinh(k\ep_2)}Q^k\right)\,,
\eq
and observe that for even charge ($+$) the summand is the same as the first integrand in \req{SchwingerInt}. This is similar as in the ordinary refined case. Furthermore, explicit expansion of the corresponding free energies (parameterized via \req{epdef}) for small $Q$ and $\lambda$ shows that \req{ConiZviaVertex} equals \req{ConiZviaProd}.  For odd charge ($-$) we infer instead that
\beq\eqlabel{Z2coniM}
\t Z^-_{U(1)}=\exp\left(\sum_{k=1}^\infty\frac{e^{\frac{1}{2}k(\ep_1+\ep_2)}(e^{k\ep_1}+e^{k\ep_2})}{k(e^{2k\ep_1}-1)(e^{2k\ep_2}-1)}Q^k\right)=\exp\left(\sum_{k=1}^\infty\frac{{e^{\frac{k}{2}(\ep_2-\ep_1)}\cosh\left(\frac{k(\ep_1-\ep_2)}{2}\right)}}{2k\sinh(k\ep_1)\sinh(k\ep_2)} Q^k\right)\,,
\eq
and this is the same as the first integral in \req{SchwingerInt2} with negative sign under an additional shift of $Q$. It is more natural to define 
$$
Z_{U(1)}^-(\ep_1,\ep_2;Q):=\t Z_{U(1)}^-(\ep_1,\ep_2;Qe^{-\frac{1}{2}(\ep_2-\ep_1)})\,.
$$ 

In particular, under this definition the relation between $Z^+_{U(1)}$ and $Z^-_{U(1)}$ is just a flip of sign of one of the $\ep_i$ combined with an overall sign change, \ie, 
\beq\eqlabel{ZpvsZmConi}
Z^+_{U(1)}(\ep_1,\ep_2;Q)=-Z^-_{U(1)}(\ep_1,-\ep_2;Q)=-Z^-_{U(1)}(-\ep_1,\ep_2;Q)\,.
\eq
Also note that 
$$
Z_{U(1)}(\ep_1,\ep_2;Q)=Z_{U(1)}^+(\ep_1,\ep_2;Q)\times  Z_{U(1)}^-(\ep_1,\ep_2;Q )\,,
$$
or, in terms of the corresponding free energies $\Fcal=\log(Z)$,
$$
\Fcal(\ep_1,\ep_2;Q)=\Fcal^+(\ep_1,\ep_2;Q)+\Fcal^-(\ep_1,\ep_2;Q )\,,
$$
due to relation \req{cossinid}. 

Hence,  it is suggested to identify $Z_{U(1)}^+$ and $Z_{U(1)}^-$ as possible partition functions of the refined topological string on $A_1\times\left(\Ocal(-1)\oplus\Ocal(-1)\rightarrow\P^1\right)$. The partition functions on $A_1\times\left(\Ocal(-2)\oplus\Ocal(0)\rightarrow\P^1\right)$ follow as usual via $Z_{U(1)}^\pm\rightarrow \left(Z_{U(1)}^\pm\right)^{-1}$. Of course, due to \req{ZpvsZmConi} there is no essential difference between $Z^+_{U(1)}$ and $Z^-_{U(1)}$. However, as we will see in the next subsection, it is useful to explicitly distinguish between the two cases.

\subsection{$U(2)$}
\label{5dU2}
Let us now discuss the pure $U(2)$ gauge theory in five dimensions, since this relates to the refined topological string on local $\P^1\times\P^1$, \ie, on $\Ocal(-2,-2)\rightarrow\P^1\times\P^1$. We denote the two K\"ahler parameters of the geometry, which correspond to the sizes of the two $\P^1$, as $t_1$ and $t_2$. We also define $Q_i:=e^{-t_i}$.

As for $U(1)$, the essential lift from four to five dimensions is implemented via the simple change of weight \req{4dto5d}. From \cite{HIV03,IKV07} one can infer that the identification between the $SU(2)\subset U(2)$ gauge theory partition function in five dimenions, $Z_{SU(2)}(\ep_1,\ep_2;a;\qfrac)$, and the refined topological string partition function (defined via M-theory left-right spin counting) on local $\P^1\times\P^1$ goes as 
$$
Z_{\P^1\times\P^1}(\ep_1,\ep_2;Q_1,Q_2)= Z_{U(1)}(\ep_1,\ep_2;Q_1e^{\frac{\ep_1+\ep_2}{2}})Z_{U(1)}(\ep_1,\ep_2;Q_1e^{-\frac{\ep_1+\ep_2}{2}}) \times Z_{SU(2)}(\ep_1,\ep_2;a;\qfrac)\,,
$$
where $Z_{U(1)}(\ep_1,\ep_2;Q)$ is as in \req{ZconiExp}, and with identification of parameters
$$
a=-\frac{1}{2}\log Q_1\,,\,\,\,\,\, \qfrac=\frac{Q_2}{Q_1} e^{\frac{1}{2}(\ep_1+\ep_2)}\,.
$$
(Usually, the two $Z_{U(1)}$ factors are referred to as perturbative contribution.) One can also interchange the role of $Q_1$ and $Q_2$, due to the symmetry of the geometry. It is interesting to note that while the full partition function $Z_{\P^1\times\P^1}$ has an expansion into even powers of $\lambda$ only, the equivariant limit to either of the $\Ocal(-2)\oplus\Ocal(0)\rightarrow\P^1$ parts of the geometry does not (only under additional shifts of the $Q$). That means that the cancellation of the odd sector in $\lambda$ is a global phenomenon.

Using the results of the previous sections, it is now easy to suggest the partition function of the refined topological string on $A_1\times\left({\rm local}~ \P^1\times\P^1 \right)$. Namely,
\beq\eqlabel{U2Zpm}
Z^\pm_{\P^1\times\P^1}(\ep_1,\ep_2;Q_1,Q_2)= Z^\pm_{U(1)}(\ep_1,\ep_2;Q_1e^{\frac{\ep_1+\ep_2}{2}})Z^\pm_{U(1)}(\ep_1,\ep_2;Q_1e^{-\frac{\ep_1+\ep_2}{2}}) \times Z^\pm_{SU(2)}(\ep_1,\ep_2;a;\qfrac)\,,
\eq
In particular the $\pm$ in  $Z^\pm_{SU(2)}$ stand for the even, respectively odd, projection discussed in section \ref{locSec}, while $Z^\pm_{U(1)}$ is given in \req{ConiZviaProd}, respectively \req{Z2coniM}. One should note that only for the same pairing of $\pm$ between the $U(1)$ and $SU(2)$ part one obtains a partition function which is symmetric under exchange of $Q_1$ and $Q_2$. Note also that while in $Z^+_{\P^1\times\P^1}$ only integer powers of $Q_i$ appear, $Z^-_{\P^1\times\P^1}$ possesses also half-integer powers, reflecting the presence of ``fractional" instantons in the effective field theory limit. In the following section, we will give strong support in favor of the interpretation of \req{U2Zpm} as a (refined) topological string partition function.

\section{B-model approach to $U(2)$ gauge theory in 5d}
\label{U25DB}
In this section we are going to reproduce the five dimensional partition functions of the previous section via the B-model approach applied to the mirror geometry of local $\P^1\times\P^1$. This hints towards the interpretation of the five-dimensional partition functions as the partition function of a refined topological string on $A_1\times$(local $\P^1\times\P^1$).

\subsection{Generalities}
The B-model formalism for the topological string on this geometry has been exhaustively discussed in the literature, hence we will be brief and only recall the necessities.  

Via mirror symmetry, the (large volume) tree-level data of the topological string on local $\P^1\times\P^1$, can be obtained by solving the system of Picard-Fuchs equations \cite{CKYZ99}
\beq\eqlabel{F0PF}
\begin{split}
\Lcal_1&=\theta^2_1-2z_1(\theta_1+\theta_2)(1+2\theta_1+2\theta_2) \,,\\
\Lcal_2&=\theta^2_2-2z_2(\theta_1+\theta_2)(1+2\theta_1+2\theta_2)\,,
\end{split}
\eq
with $\theta_i:=z_i\partial_{z_i}$ and where $z_i$ denote the two complex structure parameters mirror to the two K\"ahler parameters. The moduli space of the mirror geometry has a rich structure, see \cite{AKMV02}. However, for our purposes only the large volume point and the conifold locus are of relevance. \footnote{One could also use the point in moduli space corresponding to the geometric engineering limit to compare directly to the 4d gauge theory. We performed this analysis and obtained the expected results. However, we omit this discussion in this note for brevity, since it does not provide any major new insights.} The former to compare to the five dimensional gauge theory results, and the latter to fix the holomorphic ambiguities. In particular, since we are only interested in these points, we do not need to bother about resolving singularities in moduli space. Hence, working soley with the (singular) moduli space parameterized by the discriminant 
$$
\Delta \t\Delta =0\,,
$$
with
\beq\eqlabel{F0Delta}
\Delta=(1-8(z_1+z_2)+16(z_1-z_2)^2)\,,
\eq
and
$$
\t\Delta=\t\Delta_1\t\Delta_2\,; \,\,\,\,\, \t\Delta_i=z_i\,,
$$
is sufficient for our purposes. 

Solving the set of Picard-Fuchs equations \req{F0PF} yields the mirror maps and the prepotential $\Fcal^{(0)}$ for the large volume point ($z_1=z_2=0$). For the conifold point, we can for instance choose $z_1=z_2=\frac{1}{16}$, as this point lies only on $\Delta=0$, and is non-singular. A possible choice of coordinates around this point is (see for example \cite{HKR08})
\beq\eqlabel{F0coniL1}
\begin{split}
z^c_1=1-\frac{z_1}{z_2}\,,\,\,\,\,\, z^c_2=1-\frac{z_2}{\frac{1}{8}-z_1}\,.
\end{split}
\eq
Solving the Picard-Fuchs system \req{F0PF} in the new coordinates $z^c_i$ gives the mirror maps and the prepotential at the conifold locus. 

Having the tree-level data at the large volume and a conifold point in moduli space at hand, one can solve for the higher genus amplitudes recursively via the holomorphic anomaly equations, both, in the ordinary and refined case. Since the local $\P^1\times\P^1$ geometry effectively behaves like a 1-parameter geometry, the 1-parameter holomorphic anomaly equations introduced at hand of $SU(2)$ gauge theory in section \ref{BSU2} are sufficient. Note also that the effect of refinement is a pure change of boundary conditions at the conifold point, as anticipated in \cite{KW10a} and explicitly confirmed in \cite{HK10}.

\subsection{$Z^{+}_{SU(2)}$}

Via explicit comparision to the localization results of section \ref{5dU2}, we find that the 1-loop free energy of the lift of the gauge theory on $A_1$ to five dimensions can be reproduced in the B-model formalism via the usual solution of the 1-loop holomorphic anomaly equation 
\beq\eqlabel{F01loop}
\Gcal^{(1)}=-\frac{1}{2}\log\left(\det G\right)+a_\Gcal^{(1)}(z_1,z_2;\beta)\,,
\eq
with $G_{ij}:=\partial_{Q_i}z_j$, holomorphic ambiguity $a_\Gcal^{(1)}$ parameterized as
\beq\eqlabel{F01loopA}
a_\Gcal^{(1)}(z_1,z_2;\beta)=\kappa_1\log\Delta+\kappa_2\log\t\Delta\,,
\eq
albeit under fixing
$$
\kappa_1=-\frac{1}{48}\left(\beta+\frac{1}{\beta}+6\right)\,,\,\,\,\,\,\kappa_2=-\frac{1}{48} \left(24-\beta-\frac{1}{\beta}\right) \,.
$$
Note that the coefficient $\kappa_1$ of the ``conifold term" has the expected universal structure, \ie, is proportional to $\Psi^{(1)}_2$ given in \req{Psis}. 

Having the tree-level and 1-loop data at hand, we can try to solve for the higher genus amplitudes using the holomorphic anomaly equations supplemented with the expected boundary conditions \req{Psis} for the conifold point in moduli space. We parameterize the holomorphic ambiguity as usual, \ie, via
$$
a_\Gcal^{(g>1)}(z_1,z_2;\beta)=\frac{1}{\Delta^{2g-2}}\sum_{\substack{n_i\leq 4(g-1)}} a_{n_1,n_2}^{(g)}(\beta)\, z_1^{n_1} z_2^{n_2}\,.
$$
Similar as in the original case, the symmetry of the partition function under exchange of $z_1$ and $z_2$,  the known constant map contribution and conifold behavior yield enough constraints to fix the coefficients $a^{(g)}_{n_1,n_2}$ for arbitrary $g$. 

In this manner, the amplitudes can be calculated to high degree in $Q_i$. Clearly, the explicit results are too lengthy to be explicitly shown here. We just give the leading terms of the 2-loop amplitude for illustrational purposes, \ie, 
$$
\Gcal^{(2)}=\frac{1+40\beta^2+\beta^4}{720\beta^2}(Q_1+Q_2)- \frac{59+1440\beta-2950\beta^2+1440\beta^3+59\beta^4}{360\beta^2}Q_1Q_2+\dots\,.
$$

Finally, we also note that we have found an all integer BPS state type expansion of the five dimensional partition function. We will come back to this elsewhere.

\subsection{$Z^-_{SU(2)}$}
Let us now consider the other sector. Since in the underlying four-dimensional gauge theory the $\Z_2$ symmetry between the monopole and dyon point is broken at 1-loop, we expect also something new to happen for the refined topological B-model on $A_1\times$ (local $\P^1\times\P^1$). Indeed, the genus expansion of the free energy obtained from the localization calculation for the lifted gauge theory shows that the parameters $Q_i$ also occur with half-integer powers at 1-loop and beyond, which is not immediately clear how to be reproduced by the B-model since the ordinary 1-loop amplitude \req{F01loop} for general ambiguity \req{F01loopA} clearly has an expansion into integer powers of $Q_i$ only. However, similar as for the gauge theory in four dimensions, the solution lies in a ``refinement" of the conifold locus in moduli space at 1-loop and beyond. For that, note that the parameterization of the conifold locus ($\Delta=0$ with $\Delta$ as in \req{F0Delta}), can actually be factorized as 
$$
\Delta=\Delta_1\Delta_2\Delta_3\Delta_4=0\,,
$$
with
\beq
\begin{split}
&\Delta_1=-1+2(\sqrt{z_1}-\sqrt{z_2})\,,\,\,\,\,\Delta_2=1+2(\sqrt{z_1}-\sqrt{z_2})\,,\\
&\Delta_3=-1+2(\sqrt{z_1}+\sqrt{z_2})\,,\,\,\,\,\Delta_4=1+2(\sqrt{z_1}+\sqrt{z_2})\,.
\end{split}
\eq
As it will turn out, for our purposes it is enough to consider the two combinations $\Delta_{12}=\Delta_1\Delta_2$ and $\Delta_{34}=\Delta_3\Delta_4$. We observe that we can reproduce the localization result at 1-loop if we fix the holomorphic ambiguity of the amplitude \req{F01loop} to
$$
a_\Gcal^{(1)}(z_1,z_2;\beta)=-\frac{1}{48}\left(\beta+\frac{1}{\beta}\right)\log\Delta-\frac{1}{8}\left(\log\Delta_{12}-\log\Delta_{34}\right)-\frac{1}{48}\left(30-\beta-\frac{1}{\beta}\right)\log\t\Delta\,.
$$
Note the qualitative similarity of the first two terms to \req{U20011a1}. In particular, the breaking of the $\Z_2$ symmetry between the monopole and dyon point in the four dimensional gauge theory translates to a breaking of the degeneration of the conifold locus in moduli space. Hence, in order to obtain enough boundary conditions for the higher genus amplitudes, we have to expand the amplitudes around two different conifold points in moduli space, \ie, the conifold locus breaks into two components at 1-loop and beyond. The point corresponding to the choice of coordinates \req{F0coniL1} lies on $\Delta_3=0$, thus we take for instance as the other point $(z_1,z_2)=(\frac{1}{4},1)$, which lies on $\Delta_2=0$. We can take as coordinates around this point
$$
\t z_1^c=1-\frac{4z_1}{z_2}\,,\,\,\,\,\, \t z_2^c=1-\frac{z_2}{\frac{1}{2}+2z_1}\,.
$$
Transforming the Picard-Fuchs equations \req{F0PF} to these coordinates, one obtains as solution the corresponding mirror maps and prepotential. With the tree-level and 1-loop data at hand, we solve for the higher genus amplitudes using the holomorphic anomaly equation and expand around the large volume and the two conifold points. We observe that the gauge theory results of section \ref{5dU2} can be reproduced via parameterizing the holomorphic ambiguity via 
$$
a_\Gcal^{(g>1)}(z_1,z_2;\beta)=\frac{1}{\Delta^{2g-2}}\left(\sum_{n_i\leq 4(g-1)} a_{n_1,n_2}^{(g)}(\beta)\, z_1^{n_1} z_2^{n_2}+\sum_{\substack{n_i<8(g-1)\\n_i\,{\rm odd}}}\t a^{(g)}_{n_1,n_2} (\beta)z_1^{n_1/2}z_2^{n_2/2}\right)\,.
$$
Further, the amplitudes around the two conifold points feature two independent gaps with coefficient of the leading (singular) term $\sim\Psi_2(+\beta)$, respectively, $\sim\Psi_2(-\beta)$. This yields sufficiently many boundary conditions to solve the amplitudes to any desired order. Again, the explicit amplitudes are too lengthy to be displayed here. However, for the readers convenience we give the leading terms of the 2-loop amplitude:
\beq
\begin{split}
\Gcal^{(2)}&=\frac{(\beta-1)^2}{4}Q_1^{1/2} Q_2^{1/2}+\frac{1-50\beta^2+\beta^4}{720\beta^2}(Q_1+Q_2)\\
&-\frac{59-540\beta+1010\beta^2-540\beta+59}{360\beta^2}Q_1Q_2+\frac{5(\beta-1)^2}{2\beta^2}(Q_1^{3/2}Q_2^{1/2}+Q_1^{1/2}Q_2^{3/2})+\dots\,.
\end{split}
\eq

\section{Conclusion}
\label{conc}
In this work we have initiated the study of gauge theory on ALE space and its five-dimensional lift from a special geometry and holomorphic anomaly point of view. 

Besides having explicitly shown that the partition function of pure $SU(2)\subset U(2)$ gauge theory on the simplest example of an ALE space, namely $A_1$, can be reproduced using the ``B-model" approach of invoking special geometry and the holomorphic anomaly equation, we also showed that the partition function resulting from the naive lift to five dimensions of the gauge theory on $A_1$ still enjoys this property. In particular, the corresponding tree-level geometry is local $\P^1\times\P^1$, hinting towards that the lifted gauge theory can be identified with a sort of refined topological string on $A_1\times$(local $\P^1\times\P^1$). 

We see room for further investigations in various directions. Perhaps most interesting would be the generalization to include matter, other internal backgrounds and more general ALE space-times, in order to clarify if the $SU(2)\subset U(2)$ on $A_1$ case discussed in this note is a mere mathematical curiosity, or, if there is a general structure behind. For that, it would be useful to construct an orbifolded version of the refined topological vertex (``A-model" approach). This appears to be relatively straight-forward, \ie, the only missing essential ingredient being the proper projection of the framing factor. The application of the B-model approach seems in general to be a bit more tricky due to the observed ``refinement"  of moduli space under coupling to gravity, which is expected to be a general feature for theories on ALE space-time. It appears that one should simultaneously pursue the A- and B-model approach in order to be able to definitely fix both. It would also be beneficial to properly understand the resummation of the free energies into a generating function counting (projected) BPS-states. We have found indications that such a resummation exists. We hope to come back to these and related thematics in more detail elsewhere.

\acknowledgments

D.K. likes to thank the KITP for hospitality during the INTEGRAL11 workshop, there his interest in ALE space-times arose in a conversation with T. Okuda. The work of D.K. has been supported by a Simons fellowship, by DARPA under Grant No. HR0011-09-1-0015 and by the NSF under Grant No. PHY05-51164. The work of D.K. and D.S. has been supported by the Berkeley Center for Theoretical Physics.

\end{document}